**Implications of Mortality Displacement for Effect Modification and Selection Bias**

Honghyok Kim[1], Jong-Tae Lee[2,3], Roger D. Peng[4], Kelvin C. Fong[1,5], Michelle L. Bell[1]

[1]School of the Environment, Yale University, New Haven, CT, United States of America

[2]Interdisciplinary Program in Precision Public Health, Korea University, Seoul, Republic of Korea

[3]School of Health Policy and Management, College of Health Science, Korea University, Seoul, Republic of Korea

[4]Department of Biostatistics, Bloomberg School of Public Health, Johns Hopkins University, Baltimore, MD, United States of America

[5]Department of Earth and Environmental Sciences, Faculty of Sciences, Dalhousie University, Halifax NS B3H 4R2, Canada

Corresponding Author: Honghyok Kim

E-mail: honghyok.kim@yale.edu

Address: 301 Prospect St. New Haven, CT, United States of America.

abstract
**Abstract**

Mortality displacement is the concept that deaths are moved forward in time (e.g., a few days, several months, and years) by exposure from when they would occur without the exposure, which is common in environmental time-series studies. Using concepts of a frail population and loss of life expectancy, it is understood that mortality displacement may decrease rate ratio (RR). Such decreases are thought to be minimal or substantial depending on study populations. Environmental epidemiologists have interpreted RR considering mortality displacement. This theoretical paper reveals that mortality displacement can be formulated as a built-in selection bias of RR in Cox models due to unmeasured risk factors independent from exposure of interest, and mortality displacement can also be viewed as an effect modifier by integrating the concepts of rate and loss of life expectancy. Thus, depending on the framework through which we view bias, mortality displacement can be categorized as selection bias in the bias taxonomy of epidemiology, and simultaneously mortality displacement can be seen as an effect modifier. This dichotomy provides useful implications regarding policy, effect modification, exposure time-windows selection, and generalizability, specifically why research in epidemiology may produce unexpected and heterogeneous RR over different studies and sub-populations.

Mortality displacement is the concept that deaths are moved forward in time by exposure from when they would occur without the exposure (1-3). Researchers in environmental epidemiology traditionally have investigated mortality displacement as the "short-term" mortality displacement hypothesis (1, 4) although mortality displacement includes "long-term" mortality displacement (2, 3, 5). Under the short-term mortality displacement hypothesis, which was introduced several decades ago in air pollution–mortality epidemiology (6), environmental exposure advances deaths by only *a short time period* such as a few days. Mortality displacement is traditionally understood with the concept called frail individuals whose life expectancy is already limited and their counterpart, healthy individuals (1, 2, 7) and how loss of life expectancy of individuals is linked to variation of daily death cases (3, 8). Figure 1A shows how loss of life expectancy and frail individuals shape time-series of daily deaths. The concept of mortality displacement may be generalized to include morbidity outcomes, referred to as morbidity displacement (e.g., hospitalizations) (9), but may not be ubiquitous depending on health outcomes (e.g., low weight in birth outcomes studies). For mortality outcomes, mortality displacement always exists because "premature" mortality means the timing of death brought forward by a certain time period, and a critical research question is by how much is death advanced as this impacts the true public health burden. If the increase is largely attributable to *short-term* mortality displacement, reducing exposures will have lower public health benefits as the timing of most associated deaths would be delayed only a short time period, however, longer terms of displacement reflect larger public health burden.

In time-series analysis with Poisson regression models, mortality displacement may decrease rate ratio (RR) (10-12). More specifically, in the distributed lag model (DLM) framework, RR of every single lag variable of exposure may be decreased (Figures 1B and 1C). The summary RR of multiple lag variables, termed as cumulative RR (CRR), may be attenuated but can be precisely interpreted considering how the maximum lag period can be related to loss of life expectancy (11). Figure 1B presents daily time-series of deaths when short-term mortality displacement does not exist and how this time-series may correspond to lag-response associations and CRR depending on exposure time-windows. Figure 1C is a modification of Figure 1B when short-term mortality displacement exists, which shows that RR of multiple lags may be decreased and thereby CRR may be attenuated. CRR>1 has been interpreted as evidence against the short-term mortality displacement hypothesis (6, 11-14) (Figure 1C). Numerous studies have reported that lag-response associations have different patterns over study populations (Figure 2), which may be partly explained by sociodemographic factors, behavioral factors, and access to healthcare (15). Such different patterns may suggest that mortality displacement may differ by factors, cities, and countries (15, 16). This landscape, where CRR is decreased by mortality displacement and lag structures can vary by population, suggests that mortality displacement as a causal entity related to loss of life expectancy may be an effect modifier.

This theoretical paper shows that mortality displacement in time-series analysis can be formulated by a *built-in* selection bias of hazard ratios (HR) in Cox proportional hazard models due to unmeasured risk factors independent from exposure (17, 18). Here, we use RR including HR (19, 20). This built-in selection bias of RR is often called *depletion of susceptibles*, and *non-collapsibility* in epidemiologic and statistical literature (17, 20-27). This paper further reveals that depending on the framework through which we view bias, mortality displacement can be categorized as selection bias in the bias taxonomy of epidemiology due to unmeasured risk factors, and simultaneously mortality displacement can be seen as an effect modifier as a causal entity. Our reasoning shows that this dichotomy provides useful implications regarding policy, effect modification, exposure time-windows selection, and generalizability, specifically why research in environmental epidemiology may produce unexpected and heterogeneous RR over different studies and sub-populations.

**A Built-In Selection Bias in Cox Proportional Hazard Models**

Landmark papers in epidemiology introduced that unmeasured risk factors attenuate RR for binary exposures in Cox proportional hazard models, referred to as a built-in selection bias (17, 18). We generalize this selection bias for time-varying continuous exposures in population-based studies with the DLM framework that is a generalized version of exposure variable specifications (28-30). Consider the use of Cox models to estimate the effect of exposure on failure (e.g., death). Individual's hazard can be expressed as

$$\lambda_i(t) = Z_{i,t}\lambda_0(t)\exp(a_i(X_{i,t}, \ldots, X_{i,t-L})) \qquad \text{Eq. 1}$$

where $X_{i,t}\ldots,X_{i,t-L}$ denotes exposure variables for individual *i* from time *t* to *t-L* (*L>0*) (total *L+1* time units of cumulative exposure), which can be short-term or long-term exposures, such as air pollution exposures (31). $a_i(X_{i,t}\ldots,X_{i,t-L})$ is expressed as $\sum_{l=0}^{L}\beta_{i,l}X_{i,t-l}$. *l* denotes lag. Studies often use a moving average of *X* as a metric of cumulative exposure. This use is a special case of DLMs (32). $\lambda_0(t)$ is the common baseline hazard at *t*. $Z_{i,t}$ is a set of unmeasured risk factors independent of *X* (17, 20, 27), also called as *frailty* factor (24, 26). As such, $Z_{i,t}\lambda_0(t)$ can be seen as the heterogenous baseline risk of individuals at *t*. We assume no loss to follow-up for illustrative convenience.

The overall summary of population-averaged $a_i(X_{i,t},…,X_{i,t-L})$, termed as $a(X_{i,t},…,X_{i,t-L})$ is the logarithm of CRR for a lag period from 0 to $L$, $\hat{\bar{\beta}}(L)=\sum_{l=0}^{L}\hat{\beta}_l$. In data-analysis, this is attenuated toward the null when adjustment for $Z_{i,t}$ is not made; $\hat{\beta}_1,…,\hat{\beta}_L$ is decreased. It is also possible that $\hat{\beta}_{L+1},…,\hat{\beta}_M$ that actually equal zero are estimated incorrectly as negative values. To illustrate this, we extend the directed acyclic graph (DAG) used to explain the built-in selection bias as a collider bias in previous literature (17, 20, 26) in the DLM framework (Figure 3). At *t*, negative associational paths from $X_{t-1},…,X_{t-L},…,X_{t-M}$ to $Z$ as a set of series of autocorrelated $Z_t$ over time are open by conditioning on $Y_{t-L},…,Y_0$ (i.e., conditioning on the survivorship (17, 20, 26)). This means that survivors with higher $X_t,…,X_{t-L}$ at *t* tend to have lower $Z$ than survivors with lower $X_t,…,X_{t-L}$ due to a selection process that individuals with higher $X_u,…,X_{u-L}$ and higher $Z_u$ ($u<t$) perish faster; that is why this bias is also called *depletion of susceptibles* (18, 22, 33). As a result, an imbalance of $Z$ emerges over time and deviates $\hat{\beta}_0,…,\hat{\beta}_M$ from $\beta_0,…,\beta_M$. A negative value of $\hat{\beta}_l$s arises when the strength of an emerging negative association outweighs $\beta_l$. The strength of this bias would be higher when $Z$ in the population is higher and more heterogeneous (33). We use CRR(*M*) to denote $\exp(\hat{\bar{\beta}}(M))$ and use "actual" CRR(*M*) to denote $\exp(\bar{\beta}(M))$. We use RR$_l$ to denote $\exp(\hat{\beta}_l)$ and use "actual" RR$_l$ to denote $\exp(\beta_l)$.

**Formulating Mortality Displacement as Selection Bias**

Eq.1 generalizes the binary categorization of frail individuals and healthy individuals to a continuous spectrum of baseline risk from frail individuals to healthy individuals. Frail individuals are those with high *Z*; healthy individuals are those with low *Z*. *Z* encodes loss of life expectancy because the variation of hazard by one risk factor mathematically affects loss of life expectancy by another risk factor in multiplicative hazard models (8). Life expectancy of individuals is smaller with higher $Z$ and loss of life expectancy by $X$ is smaller with higher $Z$, which conceptually corresponds to shorter-term mortality displacement.

RR in Eq. 1 can be equivalent to that of aggregated time-series analysis with Poisson regression models in an open cohort when the probability of death at $t$ is low and $X$ is shared in the same spatial unit $s$(34) or Berkson error model for $X$ (35) is assumed,

$$\log(E[Y_{s,t}]) = a(X_{s,t}, \ldots, X_{s,t-L}) + b_s(t) + c_s(t) \qquad \text{Eq.2}$$

where $Y_{s,t}$ is the number of failures, here, deaths in $s$ at $t$, $X_{s,t}$ is the population-averaged $X_i$ or shared $X$ in $s$ and $b_s(t)$ is the population-averaged effect of $\log(Z_{i,t})$. $c_s(t)$ is the population averaged $\log(\lambda_0(t))$. In time-series studies, $a(X_{s,t}\ldots,X_{s,t-L})$ is estimated through exploiting only temporal variability of $X$ and $Y$ in $s$ (e.g., city-specific regression). Multiple spatial unit-specific estimates can be combined using pooled analysis, accounting for their statistical uncertainty. $a(X_{s,t}\ldots,X_{s,t-L})$ can also be estimated through exploiting both spatial variability and temporal variability by including the logarithm of person-time at each time interval over $s$ into Eq.2, which corresponds to Cox-equivalent Poisson models (36).

As Figure 3 indicates, an increase in $a(X_{s,t}\ldots,X_{s,t-L})$ results in more deaths of individuals with higher $Z_{i,t}$ such that $b_s(u)$ ($u>t$) decreases. $b_s(u)$ increases when $Z_u$ in some individuals increases and $a(X_{s,u}\ldots,X_{s,u-L})$ decreases. So, RR of $X_{s,t}\ldots,X_{s,t-L}$ is decreased such that CRR is attenuated unless adjustment for $b_s(t)$ is made. $b_s(t)$ would be wiggly varying over time.

Patterns of $RR_l$s (i.e., lag-response association) can be very different depending on loss of life expectancy and exposure effects including exposure time-windows (11, 15). To further show various patterns of $RR_l$s depending on $Z$ that encodes loss of life expectancy, we provide numerical examples as statistical simulations. We generated time-to-failure data for an open cohort of on average, 50000 individuals, using Eq. 1 (Table 1, Figure 5). Actual CRR(30)=$\exp(\bar{\beta}(30))$ was 0.002 and actual RR=$\exp(\beta_l)$ are shown in Figure 5 as "true". $\lambda_0(t)$ is 1. We distinguished high $Z$ and low $Z$ individuals as shown in the first two columns of Table 1 (1, 37, 38). We used Gamma distribution with mean and variance of 0.1 for $Z$ based on research on the frailty factor (21, 24, 39). Some individuals with low $Z$ at time $t$ were randomly chosen to become individuals with high $Z$ at time $t+1$, mimicking life-threatening events occurring regardless of $X$. $X$ was daily 24-hour average $PM_{10}$ level in Seoul for the period of 2006–2008 (40). We fit Poisson regression models to estimate lag-response associations and CRR(30) using aggregated time-series data constructed from simulated time-to-failure data. We added a natural cubic spline (NCS) of a variable representing time with 10 degrees of freedom (df)/year in the models as in standard time-series analysis (31) and $b_s(t)$ to show how these affect estimation. The distributions of $Z$ in Table 1 were established such that CRR(30) captures all short-term mortality displacement. CRR(30) would approximately converge to

$$CRR(30) \approx \exp(0.002 \times \frac{\text{\# of death of healthy individuals}}{\text{\# of total death}})$$

CRR(30) from Poisson regression without $b_s(t)$ generally agreed with these approximations (Table 1). Figures 5A–C show actual $RR_l$s and $RR_l$s from Poisson models. $RR_l$s from models including $b_s(t)$ were identical to actual $RR_l$s and accordingly, CRR(30) from those models was 1.002. $RR_l$s from models without $b_s(t)$ were lower than actual $RR_l$s and had different patterns depending on distribution of $Z$ (Table 1). Figure 5d shows $b_s(t)$ in one simulated sample, showing why the NCS as a standard adjustment in time-series analysis (31) cannot adjust for very wiggly $b_s(t)$.

**Mortality Displacement as Effect Modification, Not A Bias**

Environmental epidemiological literature has interpreted CRR as evidence against the short-term mortality displacement hypothesis (6, 11-14), considering frail individuals and loss of life expectancy (Figure 1). We refer to these the concept of mortality displacement in the context of frail individuals and life expectancy as the *framework of mortality displacement as effect modification*. In this framework, CRR has different causal meanings depending on exposure effects including effective exposure time-windows and *M* (11). As such, mortality displacement can be viewed as an effect modifier because CRR depends on distribution of frail individuals and healthy individuals. This framework conflicts with the concept of mortality displacement as selection bias, as shown in the previous section. We refer to this concept as the *framework of mortality displacement as selection bias*, acknowledging that the built-in selection bias of RR has been discussed in epidemiological literature (17, 18, 20-24). Table 2 summarizes how the two frameworks differently categorize mortality displacement. We illustrate why this conflict takes place and further discuss the concepts of Table 2 below.

We next discuss how causal effect is framed in general epidemiology, thereby defining RR. In epidemiology, causal effect is defined as the change of probability of *Y* by *X* and is algebraically defined as the change of *Y* in a given population by a unit increment of *X*. Although the change of probability in a given time period mathematically implies the change of life expectancy (8), this common epidemiological framing does not consider how loss of life expectancy affects $RR_l$ when $Z$ is not adjusted. Causal effect as actual RR is $E[Y_{X=1}]/E[Y_{X=0}]$ at a given time interval where $Y_{X=1}$ is $Y$ at $X=1$ in that period, and $Y_{X=0}$ is $Y$ at $X=0$ in that period; for simplicity, the denominators of the rates are assumed to be ignorable by assuming a large study population. Here, $Y$ represents the number of deaths regardless of whether they are from healthy individuals or frail individuals. Thus, actual CRR, $\exp(\bar{\beta}(M))>1$, indicates excess failures, in this case deaths, regardless of their loss of life expectancy. With knowledge of causal mechanisms, a factor or entity that deviates $E[\hat{\beta}_l]$ from $\beta_l$ is called a bias (41) such that mortality displacement is a bias.

Unlike the framework of mortality displacement as selection bias, the framework of mortality displacement as effect modification considers whether $Y$ comes from healthy individuals or frail individuals in understanding the strength of CRR (Figure 1). Although not clearly mathematically and statistically defined, this framework of mortality displacement as effect modification conceptually integrates rate and loss of life expectancy in understanding RR. For example, Figure 1C shows that CRR(7) for two effective exposure time-windows is higher than 1 and lower than CRR(3) in Figure 1B. The difference between these two CRR is the number of deaths of frail individuals resulting in short-term mortality displacement. In the previous section, CRR(30) approximation equation considers only a fraction of the number of deaths from healthy individuals, not frail individuals. Using the notations in Eq.1, we can conclude that the framework of mortality displacement as effect modification endows CRR, $\exp\left(E\left[\bar{\hat{\beta}}(M)\right]\right)>1$, with causal meaning: excess deaths with loss of life expectancy of at least approximately $M$ days (11) while this meaning depends on effective exposure time-windows (11). So, $\exp(E[\bar{\hat{\beta}}(M)])$ itself can be seen as a causal estimand instead of being a biased estimate.

This dichotomy of mortality displacement has policy-relevant implications because RR can imply differential efficacy of policies across target populations depending on their baseline risk. Reducing $Z$ as the baseline risk may synergistically delay the timing of excess outcomes attributable to $X$, thereby decreasing the number of attributable premature outcomes within a period of interest. For illustration, suppose that $X$ for one week advances the timing of deaths by only seven days (i.e., seven days of life lost by $X$) (Figure 6a). Then, reducing $X$ for one week only delays premature death of very frail individuals by one week; increments of $Y$ at time 0 will be distributed to $Y$ from time 0 to time 7 in Figure 6a. In this case, reducing $X$ will not decrease annual mortality rate in a target population (2). If risk factors of these individuals other than $X$ are reduced (i.e., reducing $Z$), the situation will change. Frail individuals will become healthier, and potential gains of life expectancy by reducing $X$ will become larger; for example, in Figure 6b, increments of $Y$ at time 0 will be distributed to $Y$ from time 0 to time 730. Then, reducing $X$ may decrease annual mortality rate as much as the number of individuals whose loss of life expectancy by $X$ ranges from 366 days to 730 days (Figure 6b).

As a corollary, lower estimated RR in population with the higher baseline risk may imply that this population would experience greater health benefits of reducing $X$ if their baseline risk is reduced. This may sound counterintuitive because the epidemiological literature often assumes that higher RR has higher health benefits of reducing $X$. This assumption is correct in the framework of mortality displacement as selection bias. Depending on how we view mortality displacement, RR can have additional policy-relevant implication.

**Effect Modification When Mortality Displacement is Seen as Selection Bias, not Effect Modification**

In the framework of mortality displacement as selection bias, mortality displacement may distort observations of effect modification in actual RR. Mortality displacement systemically differs by the risk profiles of populations encoded as $Z$. The higher decrease in $RR_l$ and higher attenuation of CRR would arise in strata with individuals that, on average, have higher baseline risk. This may have striking consequences in effect modification analysis. For example, change in actual $RR_l$ by effect modification and change in $RR_l$ by mortality displacement can be counterbalanced by each other, resulting in no difference of $RR_l$ across strata although effect modification in actual $RR_l$ exists. As an extreme case, a gradient of $RR_l$ across strata can appear to reverse if differential mortality displacement outweighs effect modification in actual $RR_l$ (15).

**Mortality Displacement as Partly Mediation**

The effect of $X$ on failure can also be expressed as $X$ increases $Z_{i,t}$ (e.g., $Z_{i,t}=Z_{\cdot,t}a_i(X_{i,t}…,X_{i,t-L})$), which conceptually corresponds to the concept that $X$ makes healthy individuals frail although this is not expressed in Eq.1: $X{\rightarrow}Z{\rightarrow}Y$ in Figure 4. This is a mediation (42). This pathway changes $Z$ of individuals and thereby mortality displacement. Mortality displacement may be a mediator partly but not exclusively because ruling out the $X{\rightarrow}Y$ pathway would be a very strong assumption. With the $X{\rightarrow}Y$ pathway, mortality displacement as the built-in selection bias cannot be ruled out.

**Mortality Displacement Regarding Exposure Time-Windows Selection**

Mortality displacement may affect variable specification and selection regarding $X$, as a source of heterogeneity of research findings and publication bias (43). $X_t, ..., X_{t-M}$ is commonly specified using distributed lags of $X$ or exposure time-window methods (e.g., unweighted/weighted moving average of $X$) according to statistical indices. Examples include selection by statistical significance, information criteria, or the size of estimates. Such posteriori variable selections are easy to apply, but may induce bias (43). The fact that patterns of $\hat{\beta}_l$s would differ by study populations with different distribution of $Z$ may relate to the performance of variable selections. For example, small $\hat{\beta}_l$s decreased by mortality displacement would be less likely to result from these methods because the detection would require large sample size. If the contribution of small positive $\hat{\beta}_l$s in $\hat{\bar{\beta}}$ is ignored by variable selections, $L$ will be *further* underestimated; note that $L$ will be already underestimated because some $\beta_l$s are estimated as $\hat{\beta}_l$s≤0 due to mortality displacement as selection bias.

Furthermore, studies sometimes selectively report only some $RR_l$ or $CRR(K)$. Such reporting is informative but may increase heterogeneity of findings because the degree of mortality displacement may differ by study populations. Choosing variable selection a priori and reporting multiple $CRR(K)$s with different $K$ values along with multiple patterns of $RR_l$s will reduce unnecessary heterogeneity of findings and clarify causal interpretability of CRR (11) as well as avoid publication bias.

It is noteworthy that $CRR(\infty)$ may be 1 because it consists of all the $RR_l$ decreased by mortality displacement. In practice, CRR with very high $K$ may not be estimable due to inadequate sample size and collinearities between lagged variables of $X$ and between the lagged variables and other measured confounders. Researchers may have to specify the maximum *K* depending on their research hypotheses.

**Discussion**

Causal knowledge is a pre-requisite for bias evaluation and interpretations of RR (41, 44). This maxim holds to understand mortality displacement and its implications. Mortality displacement may be a relevant factor contributing to heterogeneity of RR over studies and different study populations and may explain unexpected findings in effect modification analysis. RR can be identical between Cox models in cohort design, Poisson regression in time-series design, and conditional logistic regression in case-crossover design (34, 45). The concept of mortality displacement may be extended toward morbidity outcomes as morbidity displacement (9). We postulate that RR may be decreased by mortality displacement in many studies where all the time-invariant and time-varying risk factors cannot be fully measured, considering intertwined determinants from biological to socio-ecological levels with stochastic process (24, 46). We note that the implications may not be necessarily applicable to study settings where the health outcome of interest is not potentially advanced in time (e.g., birth weight), although the built-in selection bias of RR in Cox models still holds.

Due to short-term mortality displacement, the use of (C)RR from time-series studies in estimating health benefits of reducing exposure levels has been criticized and the use of (C)RR from cohort studies has been recommended (2, 7, 47). CRR can be identical across study designs (34, 45) and endowed with causal meaning (11). CRR for high *M* (e.g., year) may be estimated by time-series analysis (31). What matters in estimating health impacts of environmental exposures resulting from particular policies would be *M* rather than which study design was used to product CRR.

Appropriate interpretations of RR are critical to understand policy implications of reducing exposure. We note that the real-world benefits of reducing exposure levels are not only a function of estimated RR. In the framework of mortality displacement as effect modifier, reducing *Z*, meaning the population is healthier, increases RR and thereby increases health benefits of reducing *X*. Policymakers may be able to control *Z* in target populations through interventions and other policies. For example, our previous study found higher CRR for air pollution with cardiovascular and respiratory mortalities in communities with lower smoking prevalence (15). Lag-response associations implied that longer-term mortality displacement was associated with lower smoking prevalence (15). Community-level smoking prevalence may be related to *Z* because smoking is a risk factor of those mortalities. Given the same reduction in air pollution, reducing smoking prevalence may lead to reduced air pollution-related mortalities.

Failure to consider differential mortality displacement across subpopulation strata in analyzing and interpreting RR may lead to misunderstanding of effect modification in actual RR. In epidemiologic research, stratification and interaction analyses are commonly used to test whether, for example, lower socioeconomic status (SES), racial/ethnic minority populations, or those with poor health are more susceptible to *X*. For potentially susceptible groups, baseline risk may be, on average, high while risk may be heterogenous across individuals within these groups. Even when models adjust for subpopulation factors (e.g., SES, race/ethnicity) or when these factors are not confounders (e.g., in daily time-series analysis), *Z* is not controlled unless adjustment is made for heterogeneous baseline risk of individuals within the subpopulation, which is rarely possible in practice. In this regard, RR in susceptible populations may be more systematically decreased by mortality displacement than non-susceptible populations. Thus, differential mortality displacement across subpopulations would be a critical but often overlooked factor contributing to unexpected findings in effect modification analyses and generally, heterogenous effect estimates across studies. For example, for the health effects of environmental exposures such as air pollution and temperature, many studies showed inconsistent findings on effect modification by subpopulation such as SES (15, 48-54), pre-existing disease conditions (50, 55-57), and smoking status (15, 55, 58-60). We hypothesize that mortality displacement may partly contribute to this heterogeneity due to the presence of mortality displacement as demonstrated in many time-series studies throughout the world (11-13, 16, 61-64) and across different study designs (34, 45). Some studies showed high RR of environmental exposures in low SES and racial/ethnic minority subpopulations compared to the counterparts (51, 65). We hypothesize that these high RR may be underestimates.

Multiplicative models are common in epidemiology (e.g., Cox models, Poisson models, logistic regression). For rate differences based on additive hazard models such as Aalen's additive hazard models with continuous time scale, non-adjustment for unmeasured risk factors does not introduce bias (20). Mortality displacement in additive hazard models requires further investigation.

To isolate effect modification in actual RR from mortality displacement, adjustment for mortality displacement is needed. Although some papers addressed adjustment for time-invariant $Z$ in randomized controlled trials with time-invariant binary exposures (22, 33), research is needed for time-varying exposures over different study designs. Until such approaches are developed, we suggest an ad-hoc approach here. Considering lag-response associations (i.e., $RR_l$s), not just CRR, may be informative. Although the gradient of $RR_l$s and CRR across subpopulations may be inconsistent due to mortality displacement, some $RR_l$ may agree with anticipated findings under effect modification. We acknowledge that statistical uncertainty may be obstacles in inferring multiple $RR_l$s (43). This research area is nascent, and further research is warranted.

**Acknowledgement:** This publication was developed under Assistance Agreement No. RD835871 awarded by the U.S. Environmental Protection Agency to Yale University. It has not been formally reviewed by EPA. The views expressed in this document are solely those of the authors and do not necessarily reflect those of the Agency. EPA does not endorse any products or commercial services mentioned in this publication. Research reported in this publication was also supported by the National Institute on Minority Health and Health Disparities of the National Institutes of Health under Award Number R01MD012769. The content is solely the responsibility of the authors and does not necessarily represent the official views of the National Institutes of Health. This publication was also supported by Basic Science Research Program through the National Research of Korea (NRF) funded by the Ministry of Education (2021R1A6A3A14039711).

**Table 1. Parameters in Numerical Examples (a–c) and CRR estimated by Poisson Regression Without Adjustment for $b_s(t)$.**

| Daily population-averaged baseline failure rate for low Z individuals, $E[Z_{i,t}]$ (a) | Daily population-averaged baseline failure rate for high Z individuals, $E[Z_{i,t}]$ (b) | Proportion of high Z individuals (c) | Expected daily baseline failures from low Z individuals (d)=50000×(1-(c))×(a) | Expected daily baseline failures from high Z individuals (e) =50000×(c)×(b) | CRR(30) approximation equation $\approx \exp(0.002 \times (d)/((d)+(e)))$ | CRR(30) estimated by Poisson regression | Link with Figure 5 panels |
|---|---|---|---|---|---|---|---|
| 0.0005 | 0.25 | 0.002 | 24.950 | 25.0 | 1.00100 | 1.00109 | Figure 5a |
| 0.0005 | 0.50 | 0.001 | 24.975 | 25.0 | 1.00100 | 1.00106 | Figure 5a |
| 0.0005 | 0.75 | 0.001 | 24.975 | 37.5 | 1.00080 | 1.00087 | Figure 5a |
| 0.00025 | 0.25 | 0.002 | 12.475 | 25.0 | 1.00067 | 1.00062 | Figure 5b |
| 0.00025 | 0.50 | 0.001 | 12.488 | 25.0 | 1.00067 | 1.00066 | Figure 5b |
| 0.00025 | 0.75 | 0.001 | 12.488 | 37.5 | 1.00050 | 1.00048 | Figure 5b |
| 0.0001 | 0.25 | 0.002 | 4.9900 | 25.0 | 1.00033 | 1.00033 | Figure 5c |
| 0.0001 | 0.50 | 0.001 | 4.9950 | 25.0 | 1.00033 | 1.00034 | Figure 5c |
| 0.0001 | 0.75 | 0.001 | 4.9950 | 37.5 | 1.00024 | 1.00028 | Figure 5c |

**Table 2. Mortality Displacement as Selection Bias and Effect Modification Depending on How We Frame It.**

|  | **Selection bias** | **Effect modification** |
|---|---|---|
| Rate definition | The number of failures per population at risk within a time interval (i.e., person-time) | Rate *conceptually* integrated with loss of life expectancy. It distinguishes whether failures come from frail or healthy individuals in counting failures as CRR has been used to argue against the short-term mortality displacement hypothesis. |
| Cause of selection bias/effect modification | A set of unmeasured risk factors, $Z$, independent from $X$, as the cause of the built-in selection bias of RR in Cox proportional hazard models | Heterogeneous loss of life expectancy attributable to $X$ of individuals due to their heterogenous baseline risk encoded as $Z$ |
| Causal estimand | Actual RR=$\exp(\beta)$ or actual CRR= $\exp(\bar{\beta}(L))$ | CRR=$E[\exp(\hat{\bar{\beta}}(M))]$ |
| Degree of bias | $\exp(\beta) - E[\exp(\hat{\beta})]$ or $\exp(\bar{\beta}(L)) - E[\exp(\hat{\bar{\beta}}(L))]$ | Not seen as a bias |
| Implications regarding effect modification | Effect modification analysis may be biased due to differential mortality displacement due to heterogenous $Z$ | Causal estimand differs by distribution of $Z$ of a study population, which provides policy-relevant implications (e.g., intervening $Z$ will increase $E[\exp(\hat{\bar{\beta}}(M))]$ and thereby increase health benefits of reducing $X$ within a period of time of interest) (Figure 6) |

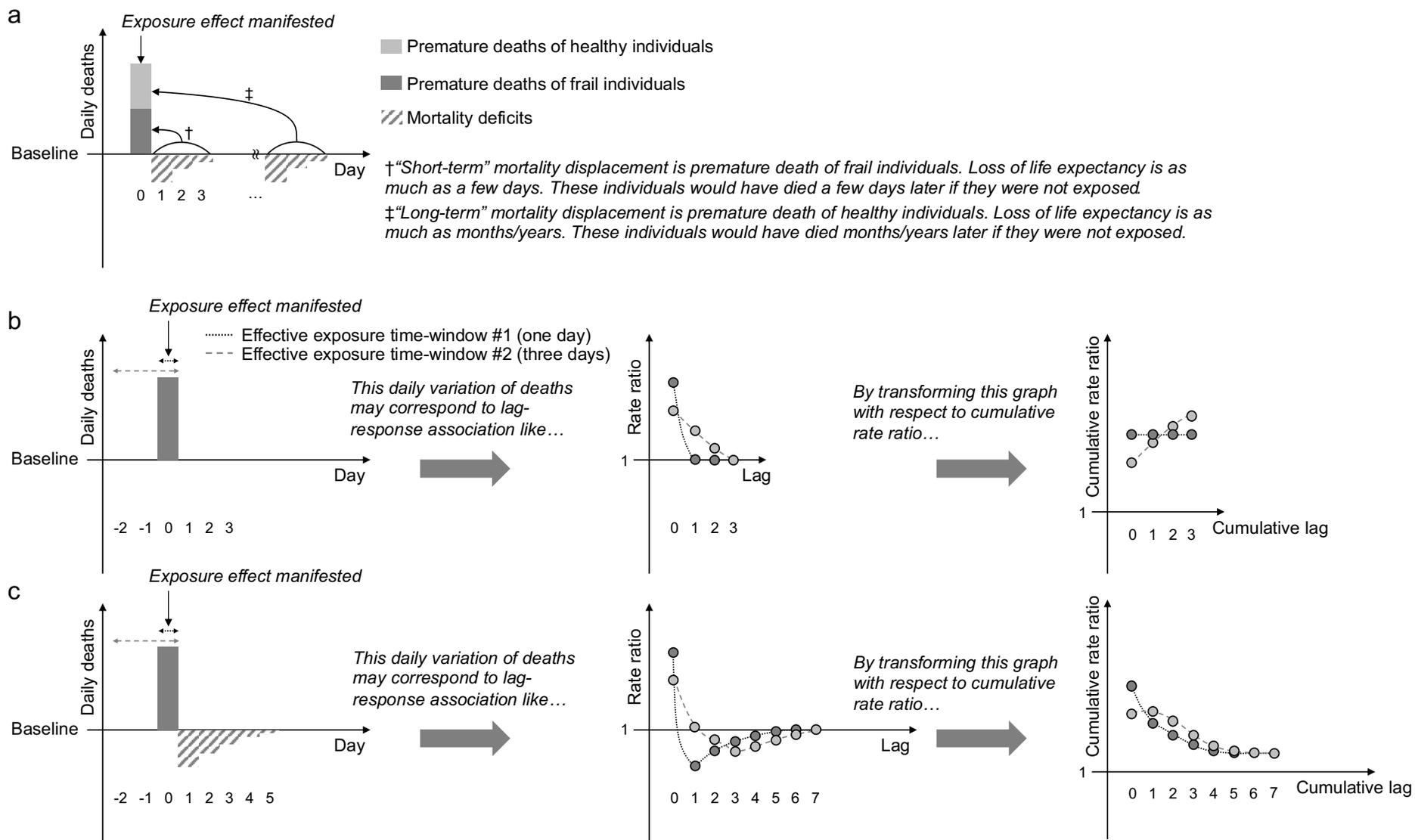

**Figure 1. Illustrations of Hypothetical Daily Time-series of Deaths, Corresponding Lag-Response Associations, and Cumulative Rate Ratios (CRR). a) Daily Time-Series of Deaths due to Short-term and Long-term Mortality Displacement. b) Daily Time-Series of Deaths, Corresponding Lag-Response Association, and CRR for Two Exposure Time-Windows When Short-Term Mortality Displacement Does Not Exist. c) Daily Time-Series of Deaths, Corresponding Lag-Response Association, and CRR for Two Exposure Time-Windows When Short-term Mortality Displacement Exists.**
Note: For readers interested in detailed illustrations about how exposure effects including exposure time-windows and mortality displacement determine lag-response associations, please see Kim and Lee (2020)'s supplementary materials (11).

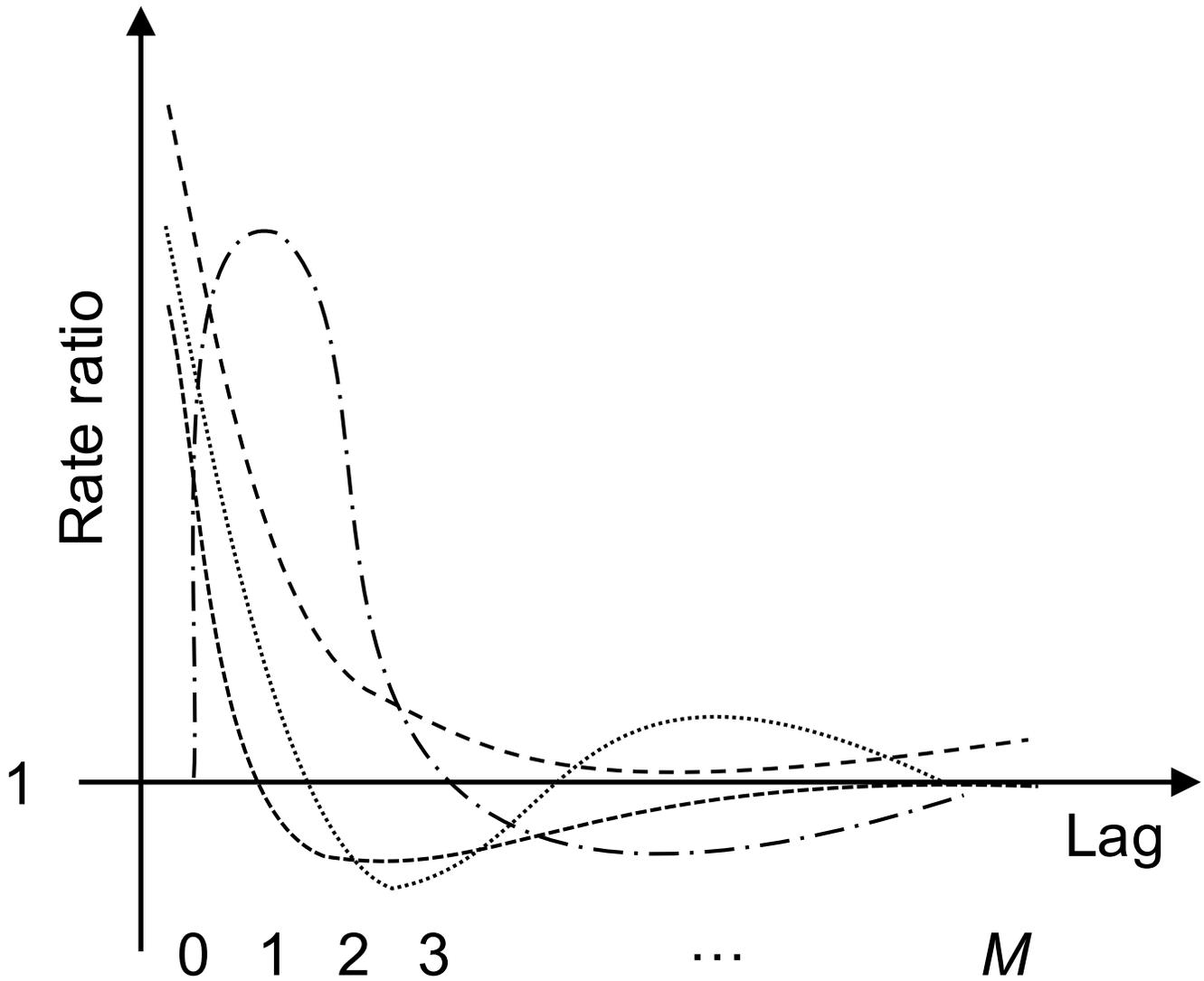

**Figure 2. Examples of Lag-Response Associations Reported in Studies for Health Effects of Air Pollution and Temperature (11-13, 15, 16, 40, 61, 66, 67).**

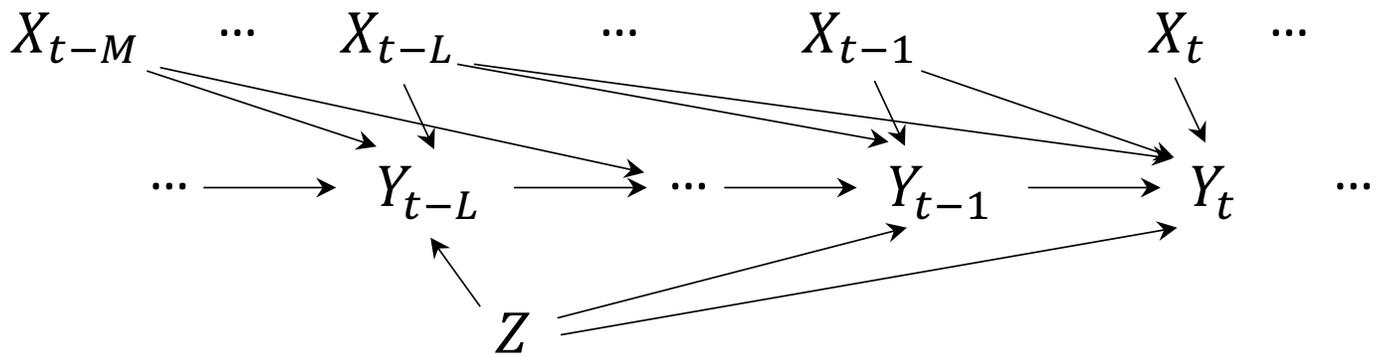

**Figure 3. Directed Acyclic Graph (DAG) for Eq. 1.**
Note: Death and survival are coded as *Y*=1 and *Y*=0, respectively. A set of unmeasured risk factors that are independent of *X* (i.e., unobserved heterogeneous baseline disease risk risk) is denoted as *Z*. For simplicity, $Z_t, \ldots, Z_{t-L}$ is denoted as *Z* because individual's baseline risk is autocorrelated over time. At time *t*, negative associational paths from $X_{t-1}, \ldots, X_{t-L}, \ldots, X_{t-M}$ to *Z* are open by conditioning on $Y_{t-L}, \ldots, Y_{t-1}$.

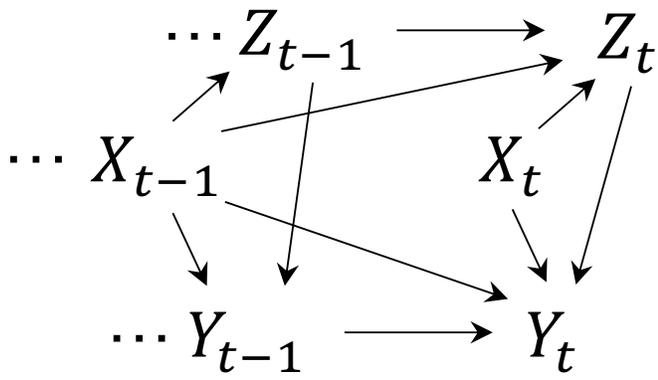

**Figure 4. Directed Acyclic Graph (DAG) for an Extended Version of Figure 3.**

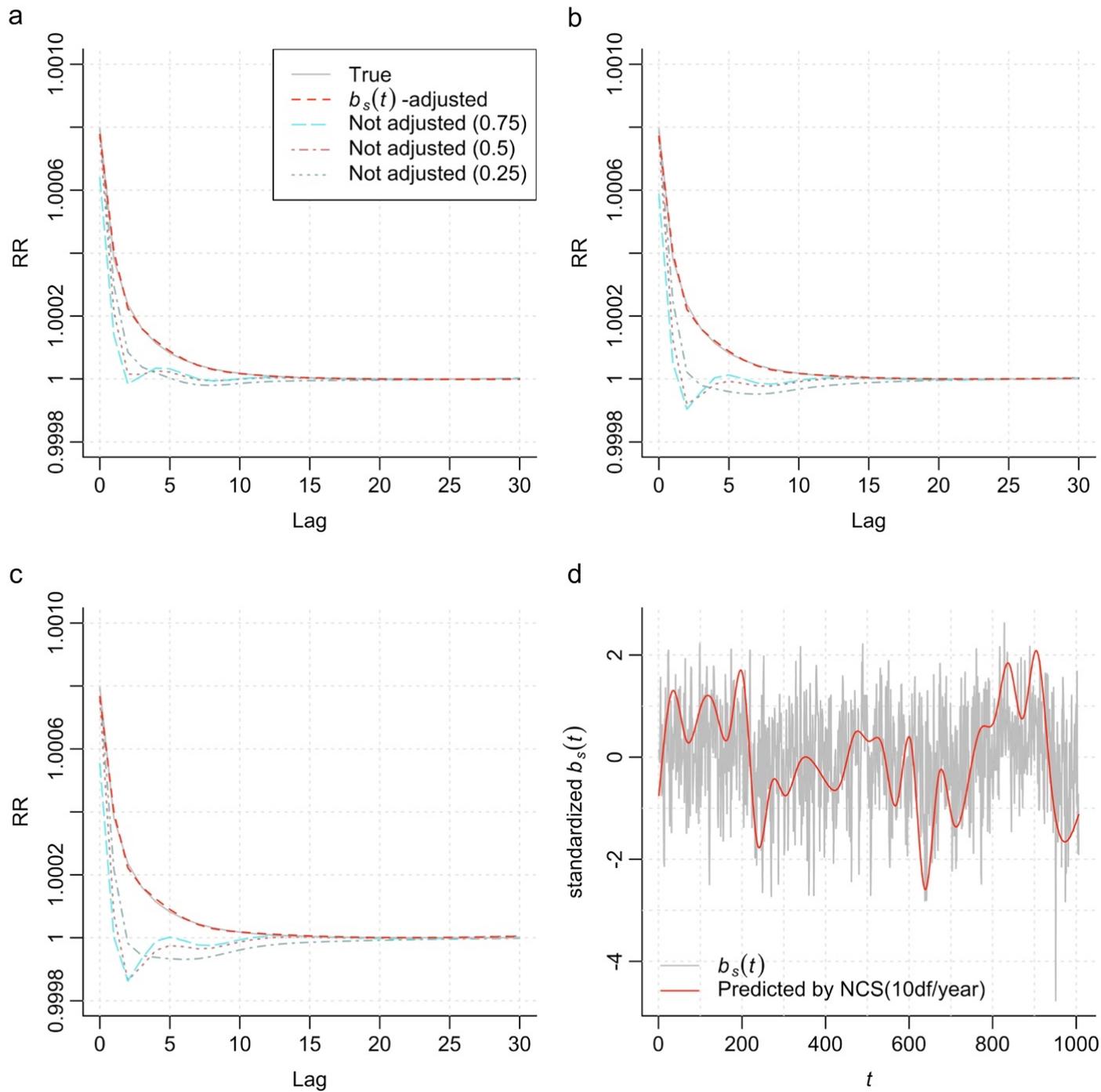

**Figure 5. Estimated Relative Rates by Aggregated Time-Series Analysis Using Poisson Regression Models With $b_s(t)$ or Without $b_s(t)$ as a Covariate. a) Daily Population-Averaged Baseline Failure Rate for Healthy Individuals Is 50 per 100,000. b) Daily Population-Averaged Baseline Failure Rate for Healthy Individuals Is 20 per 100,000. c) Daily Population-Averaged Baseline Failure Rate for Healthy Individuals is 10 per 100,000. d). Standardized $b_s(t)$ in One Simulated Sample.**
Note: The numbers in the brackets in the legend (a–c) indicates daily baseline failure rate for frail individuals.

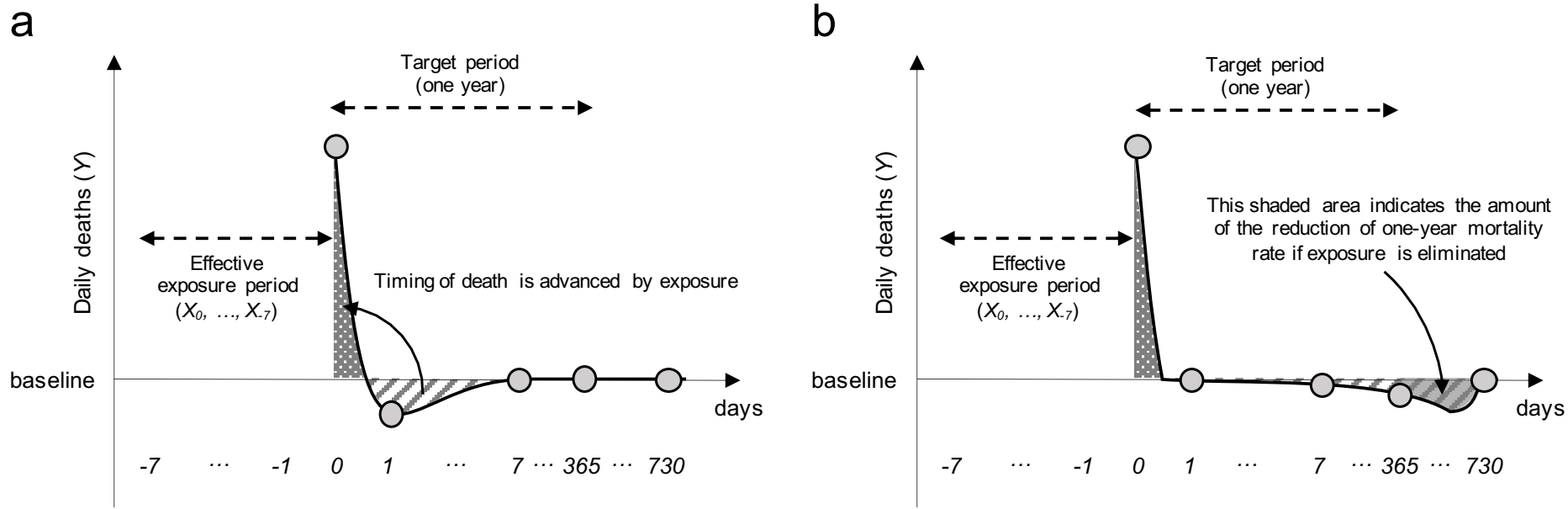

**Figure 6. Illustrations of a Causal Meaning of Mortality Displacement as Effect Modification. a) Daily Variation of *Y* when *X* for One Week (*L*=7) Increases Mortality Risk of Only Individuals Whose Potential Loss of Life Expectancy by *X* Is At Most One Week Because This Population Majorly Is Consisted of Individuals With Very High *Z*. b) Daily Variation of *Y* When *X* for One Week Increases Mortality Risk of Individuals Whose Potential Loss of Life Expectancy by *X* Ranges From 0 Day To 730 Days Because This Population Consists Largely of Individuals With Small *Z*.**